\documentclass[twocolumn,prl,citeautoscript,superscriptaddress,8pt]{revtex4-2}

\usepackage{graphicx}
\usepackage{gensymb}
\usepackage{color}
\usepackage[dvipsnames]{xcolor}
\usepackage{float}
\usepackage{upgreek}
\usepackage{bm}
\usepackage{amsmath,amssymb}
\usepackage{tabularx}
\usepackage{xfrac}
\usepackage{ulem}
\usepackage{comment}
\usepackage{makecell}

\newcommand{\VB}{\ensuremath{V_{\rm B}^{-}}}

\begin{document}

\title{Violet to near-infrared optical addressing of spin pairs in hexagonal boron nitride} 

\author{Priya Singh} % S3933111@student.rmit.edu.au
\affiliation{School of Science, RMIT University, Melbourne, VIC 3001, Australia}

\author{Islay O. Robertson} % S3930288@student.rmit.edu.au
\affiliation{School of Science, RMIT University, Melbourne, VIC 3001, Australia}

\author{Sam C. Scholten} % sam.scholten@rmit.edu.au
\affiliation{School of Science, RMIT University, Melbourne, VIC 3001, Australia}
 
\author{Alexander J. Healey} % alexander.healey2@rmit.edu.au
\affiliation{School of Science, RMIT University, Melbourne, VIC 3001, Australia}

\author{Hiroshi~Abe} % abe.hiroshi2@qst.go.jp
\affiliation{Takasaki Institute for Advanced Quantum Science, National Institutes for Quantum Science and Technology, Takasaki, Gunma 370-1292, Japan}

\author{Takeshi~Ohshima} % ohshima.takeshi@qst.go.jp
\affiliation{Takasaki Institute for Advanced Quantum Science, National Institutes for Quantum Science and Technology, Takasaki, Gunma 370-1292, Japan}
\affiliation{Department of Materials Science, Tohoku University, 6-6-02 Aramaki-Aza, Aoba-ku, Sendai 980-8579, Japan}

\author{Hark Hoe Tan} % Hoe.Tan@anu.edu.au
\affiliation{ARC Centre of Excellence for Transformative Meta-Optical Systems, Department of Electronic Materials Engineering,
Research School of Physics, The Australian National University, Canberra, ACT 2600, Australia}

\author{Mehran~Kianinia} % Mehran.Kianinia@uts.edu.au
\affiliation{School of Mathematical and Physical Sciences, University of Technology Sydney, Ultimo, NSW 2007, Australia}
\affiliation{ARC Centre of Excellence for Transformative Meta-Optical Systems, University of Technology Sydney, Ultimo, NSW 2007, Australia}

\author{Igor~Aharonovich}
%\email{igor.aharonovich@uts.edu.au}
\affiliation{School of Mathematical and Physical Sciences, University of Technology Sydney, Ultimo, NSW 2007, Australia}
\affiliation{ARC Centre of Excellence for Transformative Meta-Optical Systems, University of Technology Sydney, Ultimo, NSW 2007, Australia}

\author{David A. Broadway} % david.broadway@rmit.edu.au
\affiliation{School of Science, RMIT University, Melbourne, VIC 3001, Australia}

\author{Philipp Reineck}  
\email{philipp.reineck@rmit.edu.au}
\affiliation{School of Science, RMIT University, Melbourne, VIC 3001, Australia}
\affiliation{ARC Centre of Excellence for Nanoscale BioPhotonics, School of Science, RMIT University, Melbourne, VIC 3001, Australia}

\author{Jean-Philippe Tetienne}
\email{jean-philippe.tetienne@rmit.edu.au}
\affiliation{School of Science, RMIT University, Melbourne, VIC 3001, Australia}

\begin{abstract} 

Optically addressable solid-state spins are an important platform for practical quantum technologies. Van der Waals material hexagonal boron nitride (hBN) is a promising host as it contains a wide variety of optical emitters, but thus far observations of addressable spins have been sparse, and most of them lacked a demonstration of coherent spin control. Here we demonstrate robust optical readout of spin pairs in hBN with emission wavelengths spanning from violet to the near-infrared. We find these broadband spin pairs exist naturally in a variety of hBN samples from bulk crystals to powders to epitaxial films, and can be coherently controlled across the entire wavelength range. Furthermore, we identify the optimal wavelengths for independent readout of spin pairs and boron vacancy spin defects co-existing in the same sample. Our results establish the ubiquity of the optically addressable spin pair system in hBN across a broad parameter space, making it a versatile playground for spin-based quantum technologies. 

\end{abstract}

\maketitle 

%\section{Introduction}

Van der Waals materials offer unique opportunities for quantum technologies due to the ability to exfoliate and integrate them into complex multi-functional heterostructures and devices~\cite{Montblanch2023}.  
In particular, hexagonal boron nitride (hBN) has emerged as a front runner for photonic and spin-based quantum technologies~\cite{Turunen2022,Aharonovich2022}. With its wide bandgap ($\approx 6$\,eV), hBN hosts a rich library
of optically active defects~\cite{Tran2016,Sajid2018,Auburger2021,Jara2021}, some of which have been observed at the single defect level thereby serving as bright single-photon emitters~\cite{Shaik2021}. These optical emitters span a broad range of emission wavelengths, including ultraviolet emitters ($\approx300$\,nm~\cite{Museur2008,Bourrellier2016,Mackoit2019}), violet-blue emitters ($\approx430$\,nm, ~\cite{Shevitski2019,Fournier2021,Gale2022}), visible emitters from green to red~\cite{Tran2016b,Jungwirth2016,Stern2019,Koperski2020,Hayee2020,Tan2022,Pelliciari2024}, the boron vacancy ($\VB$) defect which emits around 800\,nm~\cite{GottschollNM2020}, and other near-infrared (NIR) emitters up to $\approx1000$\,nm~\cite{Camphausen2020}. Among these colour centres, the $\VB$ defect is the only one that had its atomic structure unambiguously identified. Moreover, $\VB$ has a paramagnetic spin-triplet ground state which can be initialised and read out optically, a property that has attracted significant interest for quantum technologies~\cite{GottschollSciAdv2021, Liu2022NatComm, GaoNatMat2022, GongNatComms2023, HealeyNP2022, Ramsay2023, Rizzato2023}. However, the $\VB$ defect is very dim and as a result it has only been observed in ensembles thus far. The other emitters are typically much brighter than $\VB$ and a subset are believed to be associated with carbon impurities~\cite{MendelsonNatMat2021}, though their exact atomic structures remain unknown. Interestingly, some of the visible emitters have been reported to possess an addressable spin doublet and are measurable as single isolated emitters~\cite{MendelsonNatMat2021, ChejanovskyNatMat2021, SternNatComms2022, GuoNatComms2023, YangACSAPN2023, Scholten2024, Patel2023,Robertson2024,Gao2024}, making them extremely promising as qubits for quantum technologies. As we showed in Ref.~\cite{Robertson2024}, these optically addressable spin systems correspond to pairs of weakly coupled electronic spins carried by two nearby point defects, one of which is optically active. However, most studies to date have focused on the yellow-orange emission region ($\approx600$-700\,nm~\cite{MendelsonNatMat2021, SternNatComms2022,YangACSAPN2023,Scholten2024, Patel2023,Robertson2024,Gao2024}), with just one report of green emitters in a nanopowder sample (540\,nm~\cite{GuoNatComms2023}) and of red emitters in an exfoliated flake (720-760\,nm~\cite{ChejanovskyNatMat2021}), and most of these reports have not shown coherent spin control. Whether this family of spin-active emitters can be reproducibly observed and coherently controlled across many different hBN samples, let alone extended towards the ultraviolet and NIR ends of the spectrum, remains unknown. Additionally, although spin-active visible emitters have been shown to co-exist with 
 $\VB$ defects~\cite{Scholten2024} when excited by the same laser, independent optical readout of two distinct spin species would open intriguing possibilities for quantum technologies.

\begin{figure*}[t]
\centering
\includegraphics[width=0.65\textwidth]{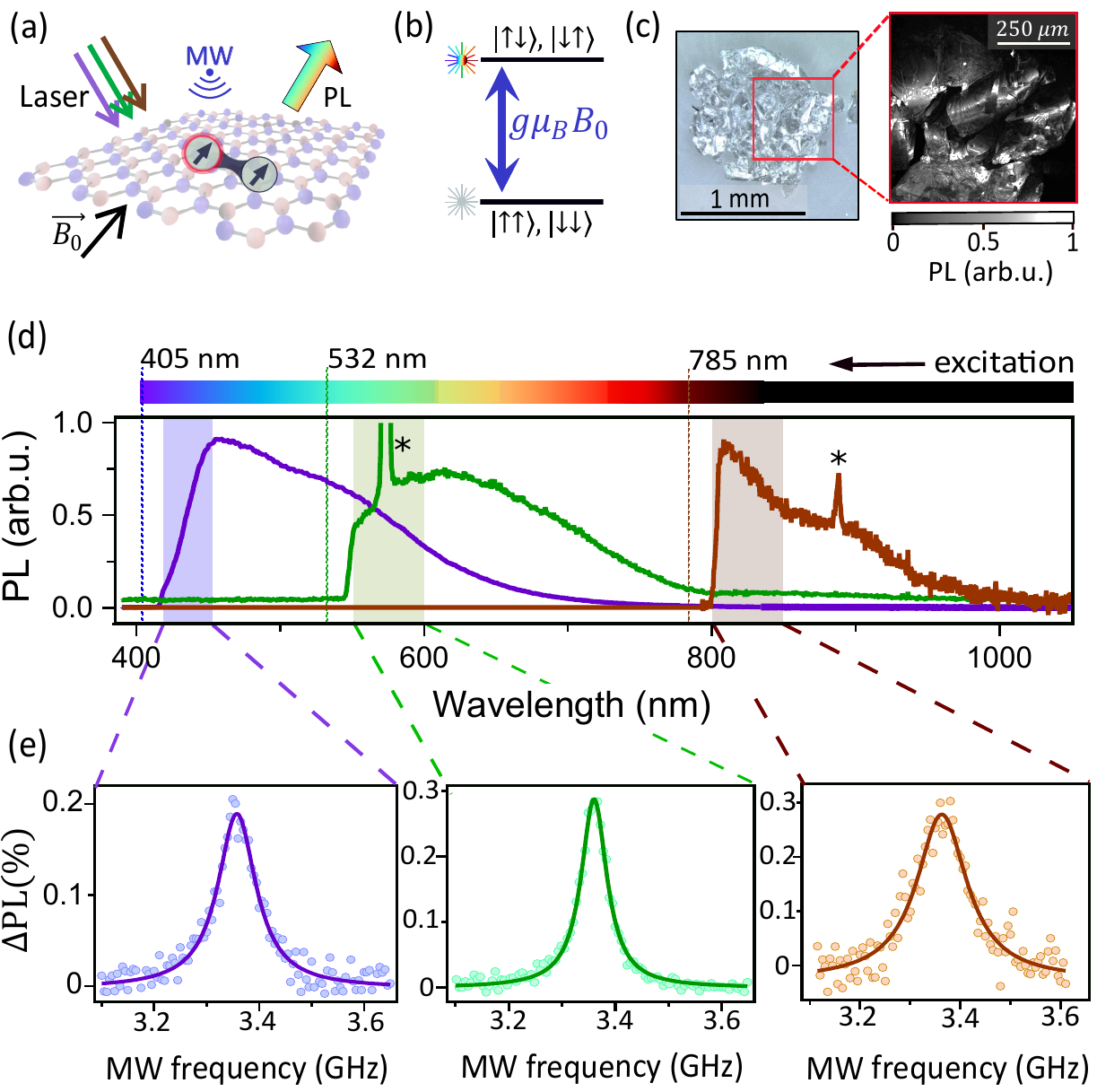}
\caption{{\bf Violet to near-infrared optical addressing of spin pairs in hBN.} 
(a)~Schematic representation of a spin pair in hBN excited by different lasers from violet to NIR. Microwave (MW) induced magnetic resonances are detected via a change in photoluminescence (PL) intensity. %the optical and spin measurements with CW 405nm, 532nm, 785nm laser excitations and readout wavelengths ranging from 405nm to 1000nm.
(b)~Simplified energy level structure of a weakly coupled spin pair under a magnetic field $B_0$. The eigenstates are grouped into two effective levels, pure triplet states $\{|{\uparrow \uparrow}\rangle,|{\downarrow \downarrow}\rangle\}$ on the one hand and singlet-triplet mixtures of $\{|{\uparrow \downarrow}\rangle,|{\downarrow \uparrow}\rangle\}$ on the other, with an energy separation of $hf_r=g\mu_BB_0$. 
(c)~ Photograph of a hBN crystal (left) and corresponding confocal PL map (collection band $\lambda_{\rm PL}=660-760$\,nm) under $635$\,nm laser illumination (right).
(d)~Ensemble-averaged PL spectra from a bright domain of the crystal under three different laser wavelengths: 405\,nm (purple line), 532\,nm (green), 785\,nm (brown). The stars (*) indicate the hBN Raman line. 
(e)~ODMR spectra of the spin pairs with an external magnetic field of $B_0\approx120$\,mT for each of the three laser excitation wavelengths. The collected PL bands (shown as shaded areas in (d)) are $\lambda_{\rm PL}=420$-450\,nm (left graph, with 405\,nm laser), $\lambda_{\rm PL}=550$-600\,nm (middle, 532\,nm laser), and $\lambda_{\rm PL}=800$-850\,nm (right, 785\,nm laser).
}
\label{fig1}
\end{figure*} 

In this paper, we show that optically addressable spin pairs exist in hBN with emission wavelengths continuously ranging from the violet (420\,nm minimum investigated) up to the NIR (1000\,nm maximum investigated). These spin pairs are universally observed in all samples investigated, and can be coherently driven regardless of wavelength. Finally, we show that $\VB$ defects created by electron irradiation co-exist with the spin pairs at all wavelengths, and their respective spin states can be independently addressed via appropriate wavelength selection. Our findings establish hBN as a uniquely versatile platform for quantum technologies based on optically addressable spins.  

\section{Results and Discussion}

The principle of the experiment is depicted in Fig.~\ref{fig1}a. A laser (wavelength $\lambda_{\rm L}$) excites an optically active point defect (circled in red) in the hBN sample which in response emits photoluminescence (PL) at wavelengths $\lambda_{\rm PL}>\lambda_{\rm L}$. In the presence of a second suitable nearby point defect, under optical pumping the system may form a metastable weakly coupled spin pair with a PL intensity that depends on the state of the spin pair~\cite{Robertson2024}, see Fig.~\ref{fig1}b. When a microwave (MW) field is applied at the resonance frequency of the spin pair, namely $f_r = g\mu_B B_0/h$ where $g\approx2$ is the Land\'e $g$-factor, $\mu_B$ the Bohr magneton, $h$ Planck's constant, and $B_0$ the applied magnetic field, the spin state gets mixed and the PL increases.

\begin{figure*}[tb!]
\centering
\includegraphics[width=0.7\linewidth]{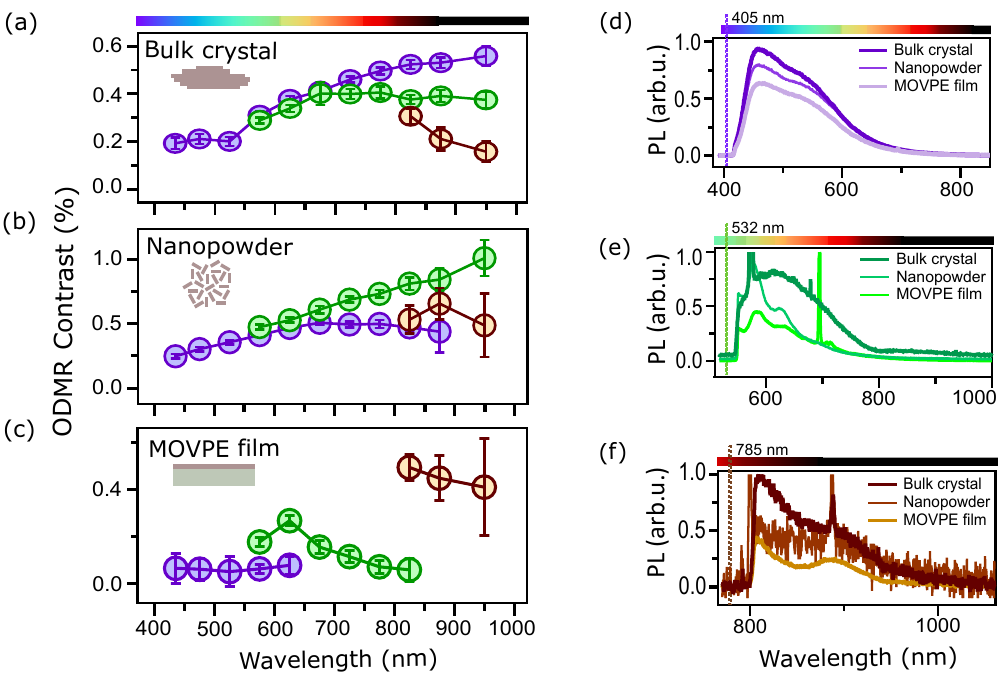}
\caption{{\bf Ubiquity of spin pairs across readout wavelengths and hBN samples.} 
(a-c)~ CW ODMR contrast as a function of PL emission wavelength for three different hBN samples (sketched in inset): (a) a bulk crystal, (b) a nanopowder, and (c) a MOVPE-grown film. For each sample, data from three different excitation lasers are shown: 405\,nm (purple markers), 532\,nm (green), and 785\,nm (brown). Each data point corresponds to a 50-nm-wide PL emission band (i.e. 450-500\,nm, 500-550\,nm, etc.) except for the left-most point (420-450\,nm band) and right-most point (900-1000\,nm band).
Error bars represent the standard error from fitting the ODMR spectrum.
(d-f)~ PL emission spectra comparing the three hBN samples for each laser wavelength: (d) 405\,nm, (e) 532\,nm, and (f) 785\,nm. 
}
\label{fig2}
\end{figure*}

We first investigate an as-received bulk hBN crystal sourced from HQ Graphene. A confocal PL image, here showing the red PL emission (660-760\,nm), reveals regions with relatively uniform PL extending over tens of microns, separated by darker regions (Fig.~\ref{fig1}c). These domains are attributed to different levels of carbon incorporation during the growth~\cite{Onodera2019}. Throughout the rest of the paper, the PL is collected from a wide ($\approx50\,\mu$m) laser spot in order to average over a large ensemble of emitters. Moreover, we employ three different excitation lasers with $\lambda_{\rm L}=405$\,nm (violet), 532\,nm (green), and 785\,nm (NIR), with the resulting PL spectra shown in Fig.~\ref{fig1}d. For each laser, the PL is primarily concentrated within 200\,nm of the laser wavelength, with a tail extending all the way up to the upper limit of our detector ($\approx1000$\,nm). Given most single emitters in hBN have been shown to have sharp but highly variable zero-phonon lines (ZPLs)~\cite{Tran2016b,Jungwirth2016,Tan2022,Pelliciari2024}, we interpret these PL spectra as being indicative of the wide distribution of ZPLs contained in the ensemble, convolved with their respective phonon side bands (PSBs)~\cite{Jara2021}. 

To test whether these emitters are associated with an addressable spin pair, we perform continuous-wave (CW) optically detected magnetic resonance (ODMR) measurements using the PL from different wavelength bands. In Fig.~\ref{fig1}e, we show ODMR spectra for PL bands close to each laser line, ensuring that ZPLs account for a large fraction of the collected PL: $\lambda_{\rm PL}=420$-450\,nm with 405\,nm laser, $\lambda_{\rm PL}=550$-600\,nm with 532\,nm laser, and $\lambda_{\rm PL}=800$-850\,nm with 785\,nm laser. In all three cases, a resonance is observed at the expected frequency $f_r = g\mu_B B_0/h\approx3.4$\,GHz ($B_0\approx120$\,mT). This result establishes the existence of optically addressable spin pairs with ZPLs from the violet-blue (420-450\,nm) to the NIR (800-850\,nm). Moreover, the similar contrast of 0.2-0.3\% for the different bands is suggestive of a universal underlying mechanism independent of emission wavelength, supporting the two-defect model of Robertson et al.~\cite{Robertson2024}. The wide range of ZPLs likely correspond to several different distinct types of emitters (with different atomic structures), although a perturbation of a single type of emitter (e.g. due to strain or nearby charges) may also contribute to the variety~\cite{Li2024}.   

The ODMR contrast observed from the same bulk hBN crystal is plotted as a function of emission wavelength (within the respective PL spectrum) in Fig.~\ref{fig2}a for the three lasers. A clear ODMR contrast is measured for all collected wavelengths between the laser line and the upper limit of our detector ($\approx1000$\,nm), including with the 405\,nm laser. This does not imply, however, that there are ZPLs at all wavelengths, as we are not able to distinguish between ZPL and PSB. For instance, the PL in the 900-1000\,nm band could be due to the PSB of emitters with a ZPL close to the 785\,nm laser line. Nevertheless, this measurement demonstrates that ODMR can be obtained from PL emission continuously spanning from 420\,nm to 1000\,nm. Note that the ODMR contrast tends to increase with emission wavelength, for instance going from 0.2\% in the blue region to 0.6\% in the NIR with the 405\,nm laser, which may indicate a higher fraction of ODMR-inactive emitters in the PL closest to the laser line. 

Importantly, we found these results to be reproducible in a wide variety of hBN samples, including bulk crystals sourced from various academic groups, commercial nano/micropowders, and films grown through vapour phase (see SI). In Fig.~\ref{fig2}b, we show the example of a hBN nanopowder sourced from Graphene Supermarket. Just like the bulk crystal, ODMR contrast is observed for all emission wavelengths from 420\,nm to 1000\,nm. Here again the contrast tends to increase with emission wavelength, reaching a maximum of 1.0\% at 900-1000\,nm under 532\,nm excitation. As another example, Fig.~\ref{fig2}c shows the case of a hBN film grown by metal-organic vapour-phase epitaxy (MOVPE)~\cite{Chugh2018,Mendelson2021}. Again, ODMR is observed for all emission wavelengths from 420\,nm to 1000\,nm across the three lasers combined, though no ODMR was detected beyond 650\,nm with the 405\,nm laser and beyond 850\,nm with the 532\,nm laser.  The PL spectra from the above three samples are remarkably similar overall, peaking at about 450\,nm with the 405\,nm laser (Fig.~\ref{fig2}d), 600\,nm with the 532\,nm laser (Fig.~\ref{fig2}e), and 820\,nm with the 785\,nm laser (Fig.~\ref{fig2}f). Regular revivals of these peaks are consistently observed which are attributed to PSBs~\cite{Mendelson2021,Jara2021,Auburger2021}. From this comparison of very different types of samples, we conclude that hBN universally contains emitters spanning from violet to the NIR, and that regardless of wavelength a fraction of these emitters is associated with an addressable spin pair. This universality lends further support to the model of Robertson et al.~\cite{Robertson2024}, which relies on the presence of a suitable nearby defect to explain ODMR rather than on properties of the primary optically active defect itself. 

\begin{figure}[tb]
\centering
\includegraphics[width=0.4\textwidth]{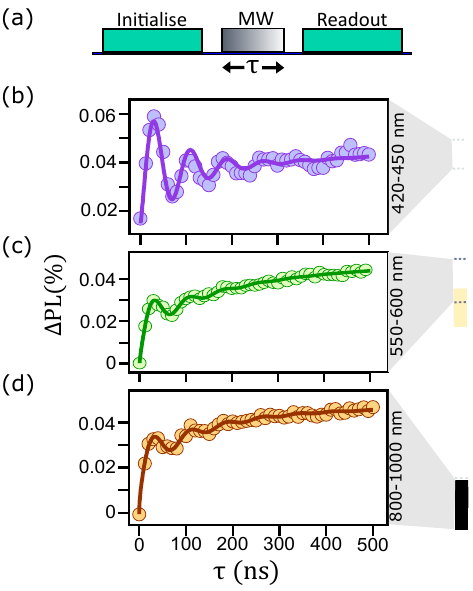}
\caption{{\bf Coherent driving of spin pairs with violet to near-infrared readout.}
(a) Pulse sequence consisting of a MW pulse of variable duration $\tau$ between two 1-$\mu$s laser pulses to initialise and read out the spin state.
(b-d)~Rabi curves acquired for a nanopowder hBN sample by collecting the PL at (b) 420-450\,nm under 405\,nm laser excitation, (c) 550-600\,nm under 532\,nm laser excitation, and (d) 800-1000\,nm under 532\,nm laser excitation.
}
\label{fig3}
\end{figure} 

Next, we test the ability to coherently drive the state of the spin pairs and optically read them out using the different readout wavelengths. For this, we apply a Rabi sequence where a laser pulse initialises the spin state, followed by a resonant MW pulse of variable duration and a second laser pulse to read out the spin state via the PL (Fig.~\ref{fig3}a). Example Rabi curves obtained are shown in Fig.~\ref{fig3}b-d using PL emitted at 420-450\,nm, 550-600\,nm, and 800-1000\,nm, respectively. In all cases, clear oscillations are resolved confirming that the spin pairs are sufficiently coherent and long-lived to sustain Rabi oscillations regardless of the emission wavelength. The more symmetric shape of the Rabi oscillations from the violet-blue PL may be an indication that the spin pairs have a longer lifetime when associated with these emitters compared to longer-wavelength emitters~\cite{Robertson2024}, motivating future work to study the wavelength-dependent photodynamics.
These results indicate that the spin pairs across all wavelengths are amenable to the application of advanced quantum control protocols~\cite{Ramsay2023,Rizzato2023}, an important pre-requisite for many applications. 

\begin{figure*}[tb]
\centering
\includegraphics[width=0.9\textwidth]{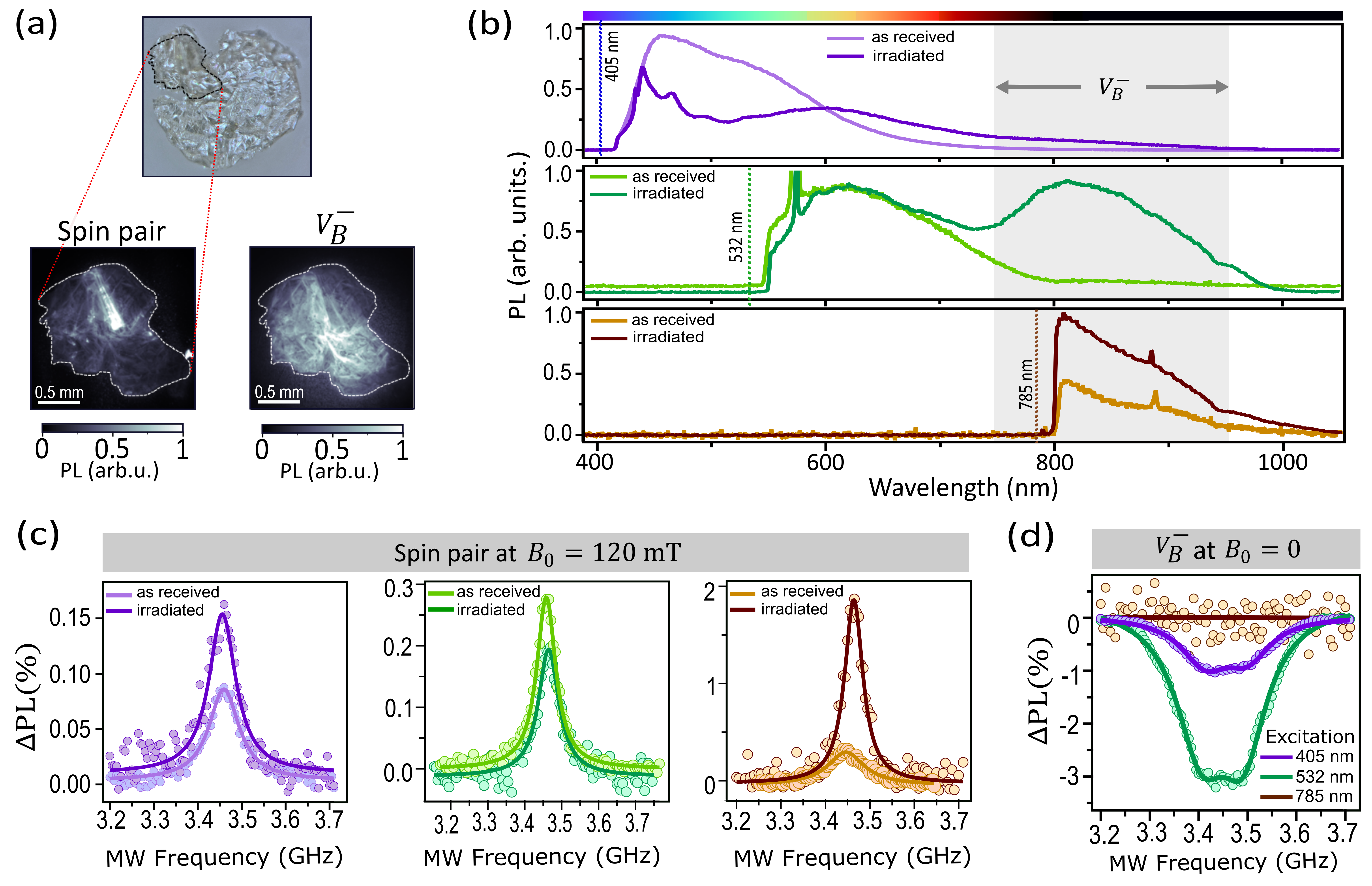}
\caption{{\bf Readout of spin pairs coexisting with $\VB$ defects.}
(a)~Photograph (top) and widefield PL images (bottom) of an electron-irradiated hBN crystal under green LED illumination, filtering for the spin pairs (bottom left, PL emission band $\lambda_{\rm PL}=600$-700\,nm), and the $\VB$ defects (bottom-right, $\lambda_{\rm PL}=850$-1000\,nm). 
(b)~PL emission spectra of the irradiated crystal from (a) compared to an as-received crystal, for three different laser excitation wavelengths: 405\,nm, 532\, nm, and 785\,nm (top to bottom). 
(c)~ODMR spectra of the spin pairs under $B_0\approx120$\,mT, comparing before and after irradiation, for each of the three excitation wavelengths. The collected PL bands are $\lambda_{\rm PL}=420$-450\,nm (left graph, with 405\,nm laser), $\lambda_{\rm PL}=550$-600\,nm (middle, 532\,nm laser), and $\lambda_{\rm PL}=800$-850\,nm (right, 785\,nm laser).
(d)~ODMR spectra of the $\VB$ defects in the irradiated crystal under $B_0=0$, for the three excitation wavelengths. The PL collection band is kept the same in all cases ($\lambda_{\rm PL}=800$-1000\,nm). 
}
\label{fig4}
\end{figure*} 

Finally, we investigate the optical readout of spin pairs at various wavelengths in hBN samples where $\VB$ defects are co-present. This is motivated by the possibility of having multiple spin species independently addressable through different lasers, which would be a useful capability for quantum technologies~\cite{Scholten2024}. To create $\VB$, a bulk hBN crystal was irradiated with 2 MeV electrons, which is expected to produce a relatively uniform density of $\VB$ throughout the crystal~\cite{Healey2024}. PL images of the irradiated crystal under green excitation reveal indeed a relatively uniform PL in the NIR attributed mainly to $\VB$, whereas the visible PL from native emitters is more patchy (Fig.~\ref{fig4}a). In Fig.~\ref{fig4}b, we compare PL spectra before and after irradiation, for the same three lasers as previously. A broad emission peak centred around 820\,nm and characteristic of $\VB$~\cite{GottschollNM2020} clearly appears after irradiation when excited with the 532\,nm laser (middle graph in Fig.~\ref{fig4}b). The presence of $\VB$ emission is much less obvious under the 405\,nm laser (top graph) which does not excite $\VB$ as efficiently~\cite{Kianinia2020}, while the 785\,nm laser (bottom graph) is above the ZPL of $\VB$~\cite{Qian2022} and so does not excite it at all. Meanwhile, the native emission is modulated by the irradiation especially in the 450-600\,nm emission band but remains present at all wavelengths, consistent with previous work~\cite{Toledo2018}.

To verify that the spin pairs can still be addressed after irradiation, we perform ODMR measurements under a magnetic field $B_0\approx120$\,mT, shown in Fig.~\ref{fig4}c for the three different lasers. The ODMR resonance of the spin pairs is essentially unaffected by the irradiation except for a change in contrast, increasing by about 50\% when collecting PL at 420-450\,nm under 405\,nm laser (left graph in Fig.~\ref{fig4}c) and decreasing slightly in the 550-600\,nm PL under 532\,nm laser (middle graph). Interestingly, when collecting the NIR PL (800-850\,nm) under 785\,nm laser (right graph), the ODMR contrast increases from 0.3\% to nearly 2\% upon irradiation, which could indicate conversion of ODMR-inactive emitters (with no suitable nearby defect~\cite{Robertson2024}) into ODMR-active ones.  
To probe the spin resonance of $\VB$, we perform ODMR under zero magnetic field and collect the PL in the 800-1000\,nm band. The resulting ODMR spectra from the irradiated crystal are shown in Fig.~\ref{fig4}d for the three lasers. As expected, the best ODMR contrast (-3\%) is achieved under 532\,nm excitation where the PL in the collected band is primarily due to $\VB$ emission~\cite{Kianinia2020}. The contrast is reduced to -1\% with the 405\,nm laser as it excites $\VB$ less efficiently. Importantly, there is no evidence of $\VB$ resonance in the ODMR spectrum under 785\,nm excitation, confirming this excitation wavelength does not excite $\VB$. These findings suggest a way to decouple the optical addressing of the two spin species, with 785\,nm exciting the NIR spin pairs (with maximal ODMR contrast) but not $\VB$, while 532\,nm is optimal for $\VB$ readout.

\section{Conclusion}

By performing PL and ODMR spectroscopy on a variety of hBN samples under various laser excitations, we show that hBN universally contains native ODMR-active optical emitters with emission wavelengths continuously spanning the 420-1000\,nm range. The different laser lines employed allow us to determine that, as a minimum, these emitters must have ZPLs in the violet-blue (405-450\,nm), green-yellow (532-600\,nm), and NIR (785-850\,nm) regions, which most likely correspond to different types of defects. The ODMR contrast is relatively consistent across emission wavelengths and between samples, supporting the interpretation that ODMR originates from a universal mechanism as we proposed previously, involving a metastable weakly coupled spin pair that forms upon charge transfer between the optical emitter and a nearby defect. The spin pairs are sufficiently coherent and long-lived to sustain Rabi oscillations regardless of the emission wavelength from violet to NIR. Finally, we showed that electron irradiation creates $\VB$ defects without significantly affecting the native ODMR-active emitters regardless of their emission wavelength (and even improving the NIR-emitting spin pairs), and identified 785\,nm as a convenient laser wavelength to selectively control the NIR spin pairs without driving the $\VB$ spins.

Our results establish the ubiquity of the optically addressable spin pair system in hBN across a uniquely broad range of readout wavelengths. This makes hBN a particularly versatile platform for spin-based quantum technologies, as the desired wavelength can be chosen to suit each application. For instance, violet emitters may find use in ODMR-based widefield magnetic imaging~\cite{Levine2019,Scholten2021} where the shorter wavelength enables an improved spatial resolution compared to existing systems such the nitrogen-vacancy centre in diamond (PL emission centred at 700\,nm)  or the $\VB$ defect in hBN (820\,nm). On the other end of the spectrum, NIR wavelengths are preferable for sensing and imaging of biological samples. Future work will seek to extend the range of wavelengths further in the NIR towards the telecom bands for quantum communication applications, and towards the ultraviolet for high-resolution imaging. The co-presence of a different spin species ($\VB$) that can be independently addressed may find utility in multi-modal or multiplexed sensing, or to investigate many-body spin physics. To advance these various applications, it will be important to progress our understanding of the electronic structure of the spin pair system, and of the atomic structure of the defects involved. This would greatly facilitate the engineering of these systems for applications, and realise the on-demand creation of single ODMR-active emitters.

%\section{Experimental Section}
\section{Methods/Experiments}

\subsection{Experimental setup}

The ensemble measurements reported in this work were conducted using a custom-built room temperature wide-field fluorescence microscope. Three different laser sources were used, with CW emission at 405\,nm (Hubner Photonics Cobolt 06-01), at 532\,nm (Laser Quantum Opus), and at 785\,nm (Thorlabs FPL785S-250), with a laser power at the sample of about 70\,mW, 200\,mW, and 25\,mW, respectively. For pulsed measurements, the CW laser emission was gated either directly or via an acousto-optic modulator. The gated laser beam was focused through a widefield lens to the back aperture of an objective lens (Nikon S Plan Fluor ELWD 20x, NA = 0.45). Photoluminescence (PL) was separated from excitation light by a dichroic beam splitter optimised for each laser wavelength, filtered using longpass and shortpass filters, and then directed to either a spectrometer (Ocean Insight Maya2000-Pro) for PL spectroscopy or a sCMOS camera (Andor Zyla 5.5-W USB3) for spin measurements, where the counts from the entire illuminated region were added together. 

The microwave (MW) signal was generated by a Windfreak SynthNV Pro signal generator, gated through a Texas Instruments TRF37T05EVM IQ modulator, and amplified using a Mini-Circuits HPA-50W-63+ amplifier. The amplified signal was fed into a printed circuit board (PCB) with a coplanar waveguide and terminated by a 50\,$\Omega$ load. The hBN samples were placed directly onto the PCB. A SpinCore PulseBlasterESR-PRO 500\,MHz pulse pattern generator controlled the gating of both the excitation laser and MW, as well as triggered the camera. The static magnetic field was applied using a permanent magnet

For acquiring the confocal PL maps of bulk crystals shown in Fig.~\ref{fig1}c and , a commercial confocal microscope (Olympus FV1200) with CW lasers at 473\,nm (15\,mW), 559\,nm (15\,mW), 635\,nm (20\,mW) was used.

\subsection{Sample preparation}

The details of the various samples (bulk crystals, powders, MOVPE film)  used in this work are given in Table~S1 of the SI.
The bulk crystal samples studied in the main text (thickness $\sim100\,\upmu$m, lateral size $\sim1$mm) were sourced from HQ Graphene. Some of these crystals were irradiated with 2 MeV electrons at a fluence of $2 \times 10^{18}\,\text{cm}^{-2}$ for Fig.~\ref{fig4} of the main text. 
Additionally, in the SI we show results from a bulk crystal supplied by NIMS, Japan.
The powder used in the main text, purchased from Graphene Supermarket (BN Ultrafine Powder), was suspended in isopropyl alcohol (IPA) at a concentration of 20 mg/mL and horn sonicated for 30 minutes. The sediment was drop-cast onto the PCB, forming a continuous film with a thickness of a few microns. A similar method was employed for other commercially sourced powders compared in Fig. S2 of the SI. Additionally, the powder studied in the main text was irradiated with 2 MeV electrons at a fluence of $1 \times 10^{18}\,\text{cm}^{-2}$ for studying the creation of $\VB$ defects, shown in Fig. S4 of the SI. 
The MOVPE film was grown on a 2-inch sapphire wafer by metal-organic vapour-phase epitaxy (MOVPE) using triethylboron (TEB) and ammonia \cite{Chugh2018}. The film, grown with a TEB flow of 30\,$\mu$mol/min, has a thickness of $\sim40$nm (determined by atomic force microscopy).

%\section*{Supporting Information}

%The Supporting Information contains PL and ODMR data from additional samples, confocal images of bulk crystals, and a discussion of the ODMR linewidth.

\section*{Acknowledgments}

The authors thank V. Ivady for useful discussions. This work was supported by the Australian Research Council (ARC) through grants CE200100010, FT200100073, FT220100053, DE200100279, DE230100192, and DP220100178, and by the Office of Naval Research Global (N62909-22-1-2028). 
The work was performed in part at the RMIT Micro Nano Research Facility (MNRF) in the Victorian Node of the Australian National Fabrication Facility (ANFF). 
The authors acknowledge the facilities, and the scientific and technical assistance of the RMIT Microscopy \& Microanalysis Facility (RMMF), a linked laboratory of Microscopy Australia, enabled by NCRIS. 
I.O.R. is supported by an Australian Government Research Training Program Scholarship.
P.R. acknowledges support through an RMIT University Vice-Chancellor’s Research Fellowship. 
Part of this study was supported by QST President's Strategic Grant ``QST International Research Initiative''.

%\section*{Conflict of Interest}

%The authors declare no conﬂict of interest.

%\section*{Data availability}

%The data supporting the findings of this study are available within the paper and its supporting information files.

%\bibliographystyle{naturemag} %naturemag, apsrev, apsrmp
\bibliography{references}

\clearpage

\onecolumngrid

\begin{center}
\textbf{\large Supporting Information}
\end{center}

\setcounter{equation}{0}
\setcounter{section}{0}
\setcounter{figure}{0}
\setcounter{table}{0}
\setcounter{page}{1}
\makeatletter
\renewcommand{\theequation}{S\arabic{equation}}
\renewcommand{\thefigure}{S\arabic{figure}}
\renewcommand{\thetable}{S\arabic{table}}

\section{Sample details}

The samples studied in this work are listed in Table \ref{tab:sample_details}, along with the figures in which they were each used. In the main text, bulk crystal 1 sourced from HQ Graphene was used, and another bulk crystal supplied by NIMS was included for comparison in Fig.~\ref{figs2}. The hBN nanopowders were sourced from Graphene Supermarket (BN Ultrafine Powder), with a specified purity of 99.0\%. Two batches of powder, nominally identical but purchased at different times, were used: `Nanopowder 1' was purchased in 2017 (used in the main text), whereas `Nanopowder 2' was purchased in 2022, included for comparison in Fig.~\ref{figs2}. Additionally, two micropowders were investigated, with different purities: Micropowder 1 (5 $\mu$m particle size, 98\% purity) was sourced from Graphene Supermarket, and Micropowder 2 (3-4 $\mu$m particle size, 99.99\% purity) was sourced from SkySpring Nanomaterials. The MOVPE film, grown with a TEB flow of 30\,$\mu$mol/min (details can be found in Ref.~\cite{Chugh2018}), has a thickness of $\sim40$nm and was identical to that studied in Ref.~\cite{Mendelson2021}.

\begin{table}[ht]
\centering
\renewcommand{\arraystretch}{1.5} 
\setlength{\tabcolsep}{6pt} 
\begin{tabular}{|c|c|c|}
\hline
\textbf{Sample} & \textbf{Irradiation} & \textbf{Figures} \\
\hline
Bulk Crystal 1  & - none & - Fig. 1, 2, 3, 4, S1, S3 \\
(HQ Graphene) & - 2 MeV electrons, dose of $5\times10^{18} \text{cm}^{-2}$  & - Fig. 4  \\
\hline
Bulk Crystal 2  & none & Fig. S2 \\
(NIMS) &  &  \\
\hline
Nanopowder 1, 99.0\% purity & - none & - Fig. 2, S2, S3, S4  \\
(Graphene Supermarket, 2017) & - 2 MeV electrons, dose of $1\times10^{18} \text{cm}^{-2}$ & - Fig. S2, S4 \\
\hline
Nanopowder 2, 99\% purity & none & Fig. S2 \\
(Graphene Supermarket, 2022) &  &  \\
\hline
Micropowder 1, 98\% purity & none  & Fig. S2 \\
(Graphene Supermarket) &  &  \\
\hline
Micropowder 2, 99.99\% purity & none & Fig. S2 \\
(Skyspring Nanomaterials) &  &  \\
\hline
MOVPE film, $\approx40\,$nm thick & none & Fig. 2, S2, S3\\
grown on sapphire \cite{Mendelson2021} &  & \\
 \hline
\end{tabular}
\caption{List of hBN samples used in this study.}
\label{tab:sample_details}
\end{table}

\section{Confocal imaging of a bulk crystal} 

\begin{figure}[t]
\centering
\includegraphics{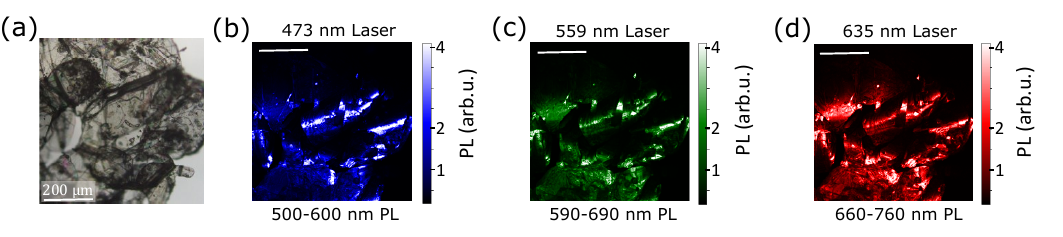}
\caption{{\bf Confocal imaging of bulk crystal.}
(a) Bright field image of the region of interest.
(b) PL map under laser illumination at 473\,nm and collection band 500-600 \,nm of the same region.
(c) Laser illumination at 559\,nm and collection band 570-670\,nm.
(d) Laser illumination at 635\,nm and collection band 660-760\,nm.
}
\label{figs1}
\end{figure}

To optically characterise the bulk crystals and evaluate the spatial homogeneity of photoluminescence (PL) across different excitation wavelengths, confocal PL imaging was performed. While Fig.~1c showed only one wavelength, here we show different illumination and collection conditions for the same region. A 0.6 $\times$ 0.6 mm region of the crystal was analysed (see bright-field image in Fig.~\ref{figs1}a) under three excitation wavelengths: 473 nm (blue), 559 nm (green), and 635 nm (red). The corresponding collected emission bands were 500-600 nm, 590-690 nm, and 660-760 nm, respectively (see Fig.~\ref{figs1}b-c), selected using appropriate filters in the commercial microscope system.

The PL images of the as-received bulk crystal reveal relatively uniform emission over tens of microns within the analyzed region, with intensities of the same order of magnitude for all three excitation wavelengths. However, direct comparison of PL intensities is challenging due to laser power variations at the focal spot, which were not measured. Spatial variations in PL were observed, with some areas exhibiting higher emission intensities in the red and blue bands compared to the green. These spatial variations could be attributed to the distribution of emitters and their associated ZPLs, as corroborated by the PL spectra shown in Fig.~1.

\section{Characterisation of additional samples} 

\begin{figure}[h]
\centering
\includegraphics[width=1\textwidth]{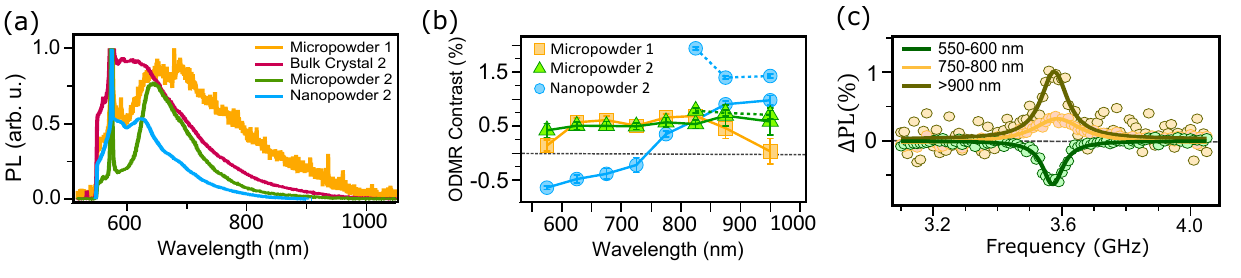}
\caption{{\bf PL and ODMR data from additional samples.}
(a) PL emission spectra of additional samples under 532\,nm laser excitationb) CW ODMR contrast as a function of PL emission wavelength for the spin pairs with 532\,nm (markers joined by solid lines) and 785\,nm (markers joined by dotted lines) illumination. Each data point corresponds to a 50-nm-wide PL emission band (i.e. 550-600\,nm, 600-650\,nm, etc.) except for the right-most point (900-1000\,nm band). For bulk crystal 2, only one ODMR measurement was performed using the 550-700\,nm PL.
(c) ODMR spectra of the spin pairs in Nanopowder 2 under $B_0\approx120$\,mT showing positive and negative contrast for different PL emission bands.
}
\label{figs2}
\end{figure}

In this section, we further analyse additional samples, specifically powder samples of varying purities sourced from different suppliers, as well as a high quality crystal from NIMS. As shown in Fig. \ref{figs2}a, the PL spectra exhibit a similar shape to the main text samples, with a peak around 600 nm under 532\,nm laser excitation. While measurement conditions vary slightly, making direct comparison of PL intensities challenging, the overall spectral similarity confirms the consistency across different samples. It can be further noted that Micropowder 1, which has lower purity compared to the other samples, shows a weaker PL relative to the other powders (as indicated by the noisier spectrum) with an emission peak centered around 650\,nm. This shift could be attributed to different emitter distributions caused by impurities in the sample.

Interestingly, when examining the ODMR contrast as a function of emission wavelength under 532\,nm excitation, a sign reversal is observed in the Nanopowder 2 sample. For this sample, the relative PL change when the MW is on resonance transitions from approximately (-0.5\%) in the 550-600\,nm emission range to about +1\% in the $>900$\,nm range with a sign flip at about 750\,nm (as shown in Fig.~\ref{figs2}b,c). Despite these variations, the ODMR effect is reproducible (and typically positive) across all samples, with ODMR detectable from PL emission ranging from 550\,nm to 1000\,nm (studied under 532\,nm and 785\,nm only). 

Bulk crystal 2 (from NIMS) is also included in Fig.~\ref{figs2}a,b, confirming that PL and ODMR can also be observed in a high purity sample. We did not perform a full wavelength-dependent study of ODMR, only collecting the PL over the 550-700\,nm range. Bulk crystals sourced from various other academic groups were also studied and gave similar results, not included here. 

\section{ODMR linewidth versus wavelength}

\begin{figure}[h]
\centering
\includegraphics[width=0.9\textwidth]{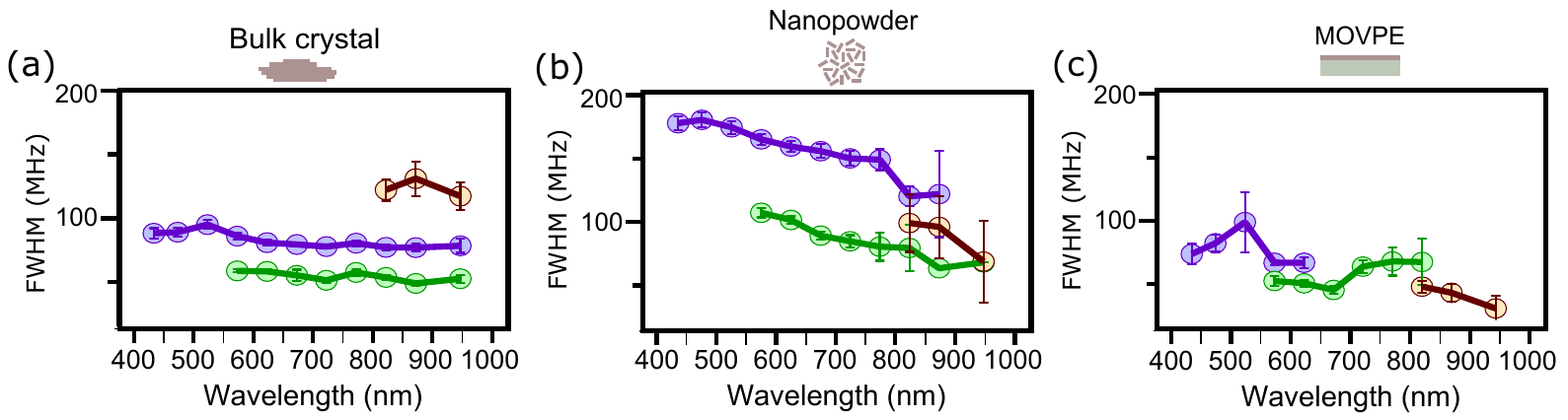}
\caption{{\bf ODMR linewidth versus emission wavelength.}
(a-c) The full width at half maximum (FWHM) extracted from Lorentzian fits of the CW ODMR spectra with respect to PL emission wavelengths for
(a) Bulk crystal 1,
(b) Nanopowder 1, 
(c) MOVPE film.
The purple, green and brown lines correspond to the 405\,nm, 532\,nm and 785\,nm laser excitations respectively. All error bars represent the standard errors from the curve fitting. Each data point corresponds to a 50-nm-wide PL emission band (i.e. 450-500\,nm, 500-550\,nm, etc.) except for the left-most point (420-450\,nm band) and right-most point (900-1000\,nm band).
}
\label{figs3}
\end{figure}

In Fig. 1e of the main text, the ODMR linewidth appears to depend on the excitation wavelength, being narrowest with 532\,nm excitation and broadest with 785\,nm excitation. To investigate this effect further, in Fig.~\ref{figs3}a we plot the ODMR linewidth as a function of PL emission wavelength for the same bulk crystal as in Fig. 1e. 
For a given laser, we observe that the linewidth (full width at half maximum, FWHM) does not vary significantly across the PL emission wavelengths, for instance with 405\,nm excitation the FWHM remains within 80-100\,MHz from the 420-450\,nm band to the 900-1000 nm band, compared to about 50\,MHz with 532\,nm excitation and 130\,MHz with 785\,nm excitation. Similar plots for the nanopowder and MOVPE film studied in Fig. 2 of the main text are shown in Fig.~\ref{figs3}b,c. There is again little variation across emission wavelengths but clear differences between laser, with here the 405\,nm laser yielding the broadest linewidth. 
This difference between lasers even for identical PL emission bands (and MW power kept constant) suggests the laser plays a dominant role in the ODMR linewidth. We leave further investigations of this effect for future work.  

\section{Irradiation of nanopowder}

\begin{figure}[t]
\centering
\includegraphics[width=0.9\textwidth]{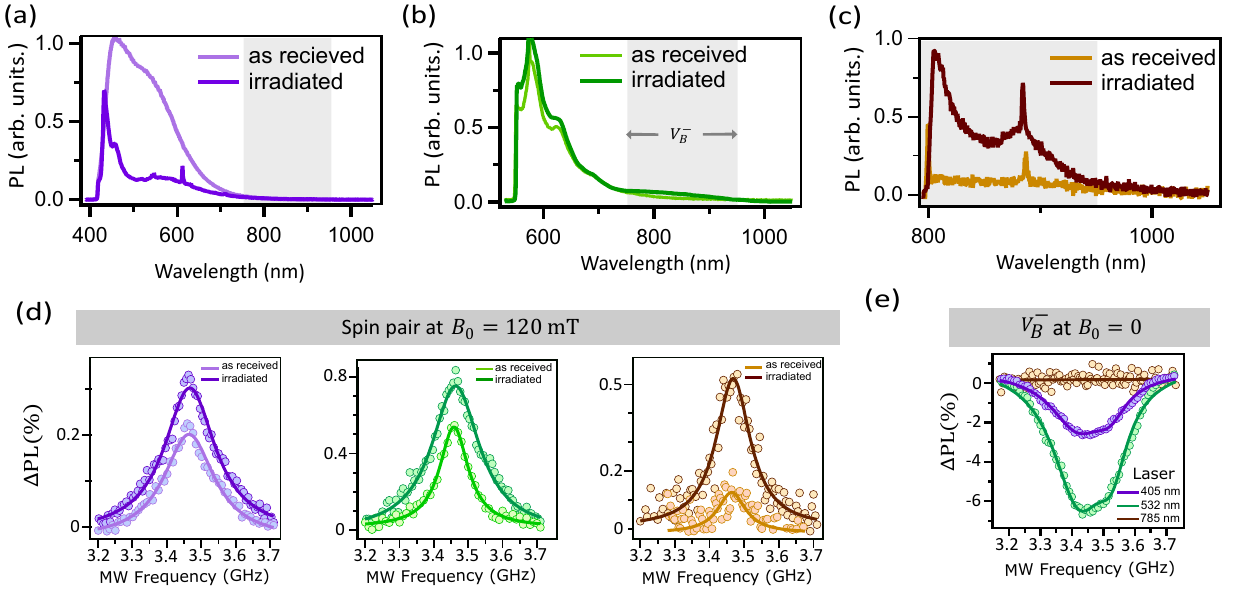}
\caption{{\bf Effect of electron irradiation in a nanopowder.}
(a-c) PL emission spectra of electron-irradiated nanopowder 1 (see table \ref{tab:sample_details}), compared to as-received nanopowder 1, for three different laser excitation wavelengths: (a) 405\,nm, (b) 532\, nm, and (c) 785\,nm. 
(d) ODMR spectra of the spin pairs under $B_0\approx120$\,mT, comparing before and after irradiation, for each of the three excitation wavelengths. The collected PL band for each laser is the same as in Fig.~4c. 
(e) ODMR spectra of the $\VB$ defects in the irradiated nanopowder 1 with the same conditions as in Fig.~4d.}
\label{figs4}
\end{figure} 

In Fig. 4 of the main text, we studied the effect of electron irradiation on a bulk crystal. Here we extend our investigation to the case of a nanopowder, and study optical readout of spin pairs where $\VB$ defects are present. To generate $\VB$ defects, the nanopowder 1 was irradiated with 2 MeV electrons, similar to the bulk crystal. However, due to the increased surface area and higher concentration of defects in the powder, we may expect a different response. First we compare PL spectra of the powder before and after irradiation across the three laser excitations in Fig.~\ref{figs4}a-c. Similar to the bulk crystal, a broad emission peak around 820\,nm under 532\,nm excitation, characteristic of $\VB$ defects, is observed but is significantly less prominent which could be due to the lower irradiation dose (by a factor~5). Under the 405\,nm (Fig.~\ref{figs4}a) and 785\,nm (Fig.~\ref{figs4}b)  the $\VB$ emission is also not apparent as expected from the bulk crystal studied in the main text. 

ODMR measurements under a magnetic field ($B_0 \approx 120$\,mT) were performed to verify the presence of the spin pairs post-irradiation. As seen in Fig.~\ref{figs4}d, the ODMR resonance of the spin pairs remains largely unaffected by the irradiation except for a change in contrast. Namely, the ODMR contrast increases slightly when collecting PL at 420-450\,nm (405\,nm laser) and 550-600\,nm (532\,nm laser), and more significantly it increases from 0.1\% to 0.5\% after irradiation when examining the NIR PL (800-850\,nm) under 785\,nm laser excitation (right graph in Fig.~\ref{figs4}d) the spin resonance of $\VB$, ODMR was performed under zero magnetic field, with PL collected in the 800-1000\,nm band. The resulting ODMR spectra for the irradiated nanopowder 1 sample are shown in Fig.~\ref{figs4}e. The 532\,nm laser again provides the strongest contrast (-6\%) due to efficient excitation of $\VB$ defects, higher than that observed in the bulk crystal (-3\%). 
Under 405\,nm excitation, the contrast is reduced to -2\%, similar to the bulk, due to less efficient excitation of $\VB$~\cite{Kianinia2020}. As expected, no $\VB$ resonance is detected under 785\,nm excitation, confirming that this wavelength does not interact with $\VB$ defects even in a nanopowder. %Overall, the powder samples demonstrate higher ODMR contrast for $\VB$ defects as well as spin pairs compared to the bulk crystal, particularly in the NIR region. 
%These results suggest that powder samples, despite their broader linewidths due to inhomogeneous broadening, may offer advantages in terms of sensitivity and spin readout for quantum magnetometry, especially when targeting $\VB$ defects.

\end{document}